\documentclass[12pt]{article}

\usepackage{amstex}
\usepackage{amssymb}
\usepackage{amscd}
\usepackage{theorem}
\usepackage{a4}

\def\calh{{\cal H}}

\def\calf{{\cal F}}

\def\Section#1{\section{#1}\setcounter{equation}{0}}

\newtheorem{proposition}{PROPOSITION}

\begin{document}

\baselineskip=0.7cm
\begin{center}
{\Large\bf GENERALIZED COHOMOLOGIES AND THE PHYSICAL SUBSPACE OF THE $SU(2)$
WZNW MODEL}
\end{center}
\vspace{0.75cm}

\begin{center} Michel DUBOIS-VIOLETTE \\
\vspace{0.3cm} {\small Laboratoire de Physique Th\'eorique et Hautes
Energies\footnote{Laboratoire associ\'e au Centre National de la Recherche
Scientifique - URA D0063}\\ Universit\'e Paris XI, B\^atiment 211\\ F-91 405
Orsay
Cedex, France\\ flad$@$qcd.th.u-psud.fr}\\
and\\
Ivan T. TODOROV\\
\vspace{0.3cm} {\small Bogoliubov Laboratory of Theoretical Physics, Joint
Institute for Nuclear Research, 141980 Dubna, Moscow Region, Russia\\
and\\
Institute for Nuclear Research and Nuclear Energy, Bulgarian Academy of
Sciences, BG-1784 Sofia, Bulgaria \footnote{Permanent address}\\

todorov$@$inrne.acad.bg}

\end{center} \vspace{1cm}

\begin{center} \today \end{center}

\vspace {1cm}

\noindent L.P.T.H.E.-ORSAY 97/08\\
\noindent BLTP-JINR, Dubna E2-97-113

\newpage
\begin{abstract}
The zero modes of the monodromy extended $SU(2)$ WZNW model give rise to a
gauge theory with a finite dimensional state space. A generalized BRS operator
$A$ such that $A^h=0\ (h=k+2=3,4,\dots$ being the height of the current algebra
representation) acts in a $(2h-1)$-dimensional indefinite metric space ${\cal
H}_I$ of quantum group invariant vectors. The generalized cohomologies
$\mbox{Ker}\  A^n/\mathrm{Im} A^{h-n}\  (n=1,\dots, h-1)$ are 1-dimensional.
Their direct sum spans the physical subquotient of ${\cal H}_I$.
\end{abstract}

\section*{Introduction}

The canonical approach to the quantum Wess-Zumino-Novikov-Witten (WZNW) model
(see \cite{AF} \cite{F} \cite{G} \cite{FG} \cite{FHT1} \cite{FHT2} \cite{FHT3}
and references therein) singles out a non-trivial finite dimensional problem
involving the (``quantum group" \cite{AF}) zero modes, its infinite dimensional
counterpart (generated by the chiral vertex operators) being relatively
trivial. The discussion in \cite{F} \cite{FG} is based on a lattice formulation
of the theory whose continuum limit is treated (following the canonical
approach of \cite{G} \cite{FG}) in \cite{FHT2}, \cite{FHT3}. The result is a
finite dimensional gauge model~: the physical state space $\calh$ appears as a
quotient, $\calh=\calh'/\calh''$, where $\calh'$ is singled out in the original
(still finite dimensional) indefinite metric space by a finite set of
constraints while $\calh''$ is the subspace of ``null" (zero norm) vectors of
$\calh'$.\\

In this paper we realize the physical space $\calh$ as a direct sum of
generalized cohomologies (of the type considered in \cite{D-VK} and \cite{D-V},
see also  \cite{Kap}). For a level $k(=1,2,\dots)$ representation of the
$A^{(1)}_1\ (=\hat{\mathfrak{su}}(2))$ Kac-Moody algebra, the ``BRS charge" $A$
satisfies $A^h=0$ where $h=k+2$ (in general, the height $h$ is the sum of the
level $k$ and the dual Coxeter number of the underlying simple Lie algebra).
The reduction of the original $h^4$ dimensional tensor product space $\calf
\otimes \bar{\calf}$ of chiral zero modes is performed in two steps. First, in
Section 1, we review the results of refs \cite{FHT2}\cite{FHT3} ending up with
a $(2h-1)$ dimensional space $\calh_I$ of $U_q(sl_2)\otimes U_q(sl_2)$
invariant vectors (with a deformation parameter $q=e^{i\frac{\pi}{h}})$.
Secondly, in Section 2,we represent $\calh$ as a direct sum of 1-dimensional
cohomologies $\mbox{Ker}\  A^n/\mathrm{Im} A^{h-n},\ n=1,\dots,h-1$ ($A^n$
being regarded as endomorphisms of $\calh_I$).

\Section{The Fock space of left and right zero modes and its quantum group
invariant subspace}

The solution of the WZNW equations of motion is customarily written in a
factorized form~:
\begin{equation}
g(x,t)=u(x-t)\bar u(x+t)\ \ (\mbox{classically,}\ \ g,u,\bar u\in SU(2)).
\label{eq1.1}
\end{equation}
Each of the quantized chiral fields, $u$ and $\bar u$, is expanded into a sum
of a positive and a negative frequency part multiplied by a ($q$-deformed)
creation and annihilation operator~:
\begin{equation}
\begin{array}{cc}
u^\alpha_\beta(x)=a^+_\beta u^\alpha_+(x,N)+a^-_\beta u^\alpha_-(x,N),\bar
u^\beta_\gamma(\bar x)=\bar a^\beta_+\bar u^+_\gamma(\bar x,\bar N)+\bar
a^\beta_-\bar u^-_\gamma(\bar x,\bar N),\\
\alpha,\beta,\gamma=1,2.
\end{array}
\label{eq1.2}
\end{equation}
Here $N$ and $\bar N$ (defined modulo $2h$ for $h=k+2$) are the (quantum)
dimension operators~:
\begin{equation}
q^N a^{\pm}=a^\pm q^{N\pm 1},\ q^{\bar N}\bar a_\pm=\bar a_{\pm}q^{\bar N\pm
1},\ \ q^h=-1,([2]:=)q+q^{-1}=2\cos \frac{\pi}{h}.
\label{eq1.3}
\end{equation}
The Fock space vacuum $\vert {vac}\ \rangle$ is assumed to satisfy
\begin{equation}
a^-\vert {vac}\ \rangle=0=\bar a_-\vert {vac}\ \rangle, \ \ q^{N-1}\vert {vac}\
\rangle=\vert {vac}\ \rangle=q^{\bar N-1}\vert {vac}\ \rangle.
\label{eq1.4}
\end{equation}
Introducing the left and right monodromies $M$ and $\bar M$ by
\begin{equation}
u(x+2\pi)=u(x)M,\ \ \bar u(\bar x+2\pi)=\bar M^{-1}\bar u (x)
\label{eq1.5}
\end{equation}
(and viewing $M$ and $\bar M$ as new dynamical variables) we demand that the
expansion (\ref{eq1.2}) diagonalizes the monodromy matrices~:
\begin{equation}
a^\pm M=a^\pm q^{\mp(N+\frac{1}{2})},\ \ \bar M \bar a_\pm =q^{\mp(\bar
N+\frac{1}{2})}\bar a_\pm.
\label{eq1.52}
\end{equation}
Furthermore, $M$ and $\bar M$ commute with the chiral vertex operators $u_\pm$
and $\bar u^\pm$. The quantum exchange relations for $u$ and $\bar u$ (derived
within the canonical approach in \cite{FG} and \cite{FHT1}) imply the following
quadratic relations among the ``$q$-oscillators"~:
\begin{equation}
\begin{array}{ll}
a^\varepsilon a^\varepsilon \prod_A (=a^\varepsilon_\rho a^\varepsilon_\sigma
(\prod_A)^{\rho\sigma}_{\alpha\beta})=0\ \  (\varepsilon=\pm),\  a^\mp a^\pm
\prod_A=\pm \frac{[N\pm 1]}{[2]}{\cal E},\\
\\
(a^-a^+-a^+a^-)\prod_A=a^-a^+-a^+a^-=[N]{\cal E}, \
[N]=\frac{q^N-q^{-N}}{q-q^{-1}}
\end{array}
\label{eq1.6}
\end{equation}
and identical relations for the bar sector $(a^\pm\rightarrow \bar a_\pm, \
N\rightarrow \bar N)$.\\
Here ${\cal E}(=({\cal E}_{\alpha\beta}))$ is the $U_q(sl_2)$ invariant
($q$-skewsymmetric) tensor and $\prod_A$ is the projector on the
$q$-skewsymmetric part of the tensor product of two $U_q(sl_2)$ spinors~:
\begin{equation}
\begin{array}{c}
\left({\cal E}_{\alpha\beta}\right)= \left(\begin{array}{cc}
0 & -q^{1/2}\\
q^{-1/2} & 0
\end{array}
\right) = \left({\cal E}^{\alpha\beta}\right),\\
\\
(\prod_A)^{\alpha\beta}_{\alpha'\beta'}=\frac{1}{[2]}{\cal
E}^{\alpha\beta}{\cal E}_{\alpha'\beta'}\ \ (\prod^2_A=\prod_A,{\cal
E}\prod_A={\cal E}).
\end{array}
\label{eq1.7}
\end{equation}
On the other hand the left and right sectors completely decouple, so that
\begin{equation}
[a^\varepsilon_\beta,\bar a^{\beta'}_{\varepsilon'}]=0\ \
(=[M^\alpha_\beta,\bar M^{\alpha'}_{\beta'}]).
\label{eq1.8}
\end{equation}
The quantum generators are read off the components $M_\pm$ of the Gauss
decomposition of the monodromy matrix~:
\begin{equation}
\begin{array}{cc}
q^{3/2}M=M_+M_-^{-1},\ M_+=\left(\begin{array}{cc}
q^{-\frac{H}{2}} & (q^{-1}-q)F q^{\frac{H}{2}}\\
0 & q^{H/2}
\end{array}
\right),\\
\\
 M^{-1}_-=\left(\begin{array}{cc}
q^{-\frac{H}{2}} & 0\\
(q^{-1}-q)Eq^{\frac{H}{2}} & q^{H/2}
\end{array}
\right)
\end{array}
\label{eq1.9}
\end{equation}
The $U_q$ covariance of $a^\pm$ is expressed by the canonical exchange
relations
\begin{equation}
\stackrel{1}{M_\pm}\stackrel{2}{a^\varepsilon}=
\stackrel{2}{a^\varepsilon}R^\pm \stackrel{1}{M_\pm},\ \   \stackrel{1}{\bar
M_\pm^{-1}} R^\pm \stackrel{2}{\bar a_\varepsilon}=\stackrel{2}{\bar
a_\varepsilon} \bar M^{-1}_{\pm}
\label{eq1.10a}
\end{equation}
or
\begin{equation}
\begin{array}{c}
[E,a^\pm_1]=0=Fa^\pm_2-q a^\pm_2 F,\ [E,a^\pm_2]=a^\pm_1 q^H,\\
\\
Fa^\pm_1-q^{-1} a^\pm_1 F=a^\pm_2,\ q^Ha^\pm_1=a^\pm_1 q^{H+1},\
q^Ha^\pm_2=a^\pm_2 q^{H-1}.
\end{array}
\label{eq1.10b}
\end{equation}
The $U_q(sl_2)$ raising and lowering operators $E$ and $F$ can be expressed as
bilinear combinations of $a^+$ and $a^-$ with coefficients in the Cartan
subalgebra (and similarly for $\bar E$ and $\bar F$)~:

\begin{equation}
E=-q^{-1/2}a^+_1 a^-_1,\  F=a^+_2 a^-_2 q^{3/2-H},\  \bar E=-\bar a^2_+\bar
a^2_- q^{\frac{1}{2}+\bar H},\  \bar F=q^{\frac{1}{2}}\bar a^1_+ \bar a^1_-.
\label{eq1.11}
\end{equation}

(Note that the role of the indices 1 and 2 for the bar sector are reversed; we
have, in particular, $q^{\bar H}\bar a^1_\varepsilon=\bar a^1_\varepsilon
q^{\bar H-1}$, etc. instead of (\ref{eq1.10b}).) \\
We also note the relations
\begin{equation}
(q^{3/2}-q^{-1/2})a^\pm_1 a^\mp_2=q^H-q^{1\mp N},\ (q^{1/2}-
q^{-3/2})a^\pm_2a^\mp_1=q^H-q^{\pm N-1}
\label{eq1.12}
\end{equation}
\begin{equation}
(q^{3/2}-q^{-1/2})\bar a_\pm^1 \bar a_\mp^2=q^{-\bar H}-q^{1\mp \bar N},\
(q^{1/2}-q^{-3/2})\bar a_\pm^2 \bar a_\mp^1= q^{-\bar H}-q^{\pm \bar N-1}.
\label{eq1.13}
\end{equation}
The diagonal action of the left and right $U_q(sl_2)$ is generated by the
coproduct
\begin{equation}
L_\pm=\bar M^{-1}_\pm M_\pm \ \mbox{satisfying} \ \ [L_\pm,a^\varepsilon_\beta
\bar a^\beta_{\varepsilon'}]=0.
\label{eq1.14}
\end{equation}
In other words, we have (assuming $[X,\bar Y]=0$ for $X,Y=E,F,H$ and using the
same notation for the $U_q$ generators in the arguments of $\Delta$ as for the
operators in the first term of the tensor product in the right-hand side)
\begin{equation}
\Delta(E)=q^H\bar E+E,\  \Delta(F)=Fq^{-\bar H}+\bar F,\ \
\Delta(q^H)=q^{H+\bar H}.
\label{eq1.15}
\end{equation}
The algebra ${\cal B}\otimes \bar{\cal B}$ generated by $a^\varepsilon$, $\bar
a_\varepsilon$ (and the Cartan's) has, for $q$ satisfying (\ref{eq1.3}), a huge
ideal which will be represented by the zero operator in the vacuum Fock space
${\cal F} \otimes \bar{\cal F}$ (see \cite{HPT})~:
\begin{equation}
(a^\varepsilon_\beta)^h{\cal F} \otimes \bar{\cal F}=0=(\bar
a^\beta_\varepsilon)^h {\cal F}\otimes \bar {\cal
F}=(q^{2h\stackrel{(-)}{H}}-1){\cal F}\otimes \bar {\cal F}.
\label{eq1.16}
\end{equation}
Each of the factors ${\cal F}$ and $\bar {\cal F}$ is then an $h^2$ dimensional
space with basis ($\vert n_1,n_2\rangle$) and
($\vert \bar n_1,\bar n_2\rangle $) where
\begin{equation}
\vert n_1n_2\rangle=(a^+_1)^{n_1}(a^+_2)^{n_2}\vert {vac}\ \rangle \in {\cal
F},\ \  \vert \bar n_1\bar n_2\rangle=(\bar a^1_+)^{\bar n_1}(\bar a^2_+)^{\bar
n_2}\vert {vac}\ \rangle \in \bar {\cal F}
\label{eq1.17}
\end{equation}
$0\leq n_1, n_2, \bar n_1, \bar n_2 \leq h-1$.\\

\subsection*{Remark}

The nilpotency relation (\ref{eq1.16}) allows to define a finite dimensional
counterpart of the Bernard-Felder cohomology \cite{BF}. To this end we define a
``BRS charge" $Q_n$, with domain the (homogeneous) subspace ${\cal F}_{h+n}$ of
${\cal F}$ spanned by vectors of the form (\ref{eq1.17}) with $n_1+n_2+1=h+n$,
setting
\[
Q_n=(a^-_1 a^-_2)^n:{\cal F}_{h+n}\rightarrow {\cal F}_{h-n}
\]
while extending its action to ${\cal F}_{h-n}$ as
\[
(a^-_1a^-_2)^{h-n}:{\cal F}_{h-n} \rightarrow 0.
\]

\vspace{5mm}

Each of the associative algebras ${\cal B}$ and $\bar{\cal B}$ (and hence,
their tensor product) admits a linear anti-involution (``transposistion")
$X\rightarrow {^tX}$ and an associated $U_q(sl_2)$ invariant bilinear form
$\langle \bullet \vert \bullet \rangle$ such that
\begin{equation}
^ta^+_\alpha=-{\cal E}^{\alpha\beta}a^-_\beta,\ ^ta^-_\alpha={\cal
E}^{\alpha\beta}a^+_\beta,\  ^t\bar a^\beta_+=\bar a^\alpha_-{\cal
E}_{\alpha\beta},\ ^t\bar a^\beta_-=-\bar a_+^\alpha {\cal E}_{\alpha\beta}
\label{eq1.18}
\end{equation}
and
\begin{equation}
\langle m_1m_2\vert n_1n_2\rangle = q^{-n_1n_2}[n_1]![n_2]!\delta_{m_1
n_1}\delta_{m_2 n_2}.
\label{eq1.19}
\end{equation}
The following statement is a consequence of Proposition 4.1 of \cite{FHT3}.
\begin{proposition}

The operators $\Delta(E), \Delta(F)$ (\ref{eq1.15}) and
\begin{equation}
B=a^+_\alpha \bar a^\alpha_-,\ ^tB=-a^-_\alpha \bar a^\alpha_+
\label{eq1.20}
\end{equation}
give rise to two commuting copies of $U_q(sl_2)$~: $[^{(t)}B,\Delta(X)]=0$ for
$X=E,F$ and
\begin{equation}
\begin{array}{c}
[\Delta(E),\Delta(F)]=[H+\bar H] =\Delta([H]),\  q^{H+\bar
H}\left(\begin{array}{c}
\Delta(E)\\
\Delta(F)
\end{array}
\right)
= \left(\begin{array}{c}
\Delta(E)\\
\Delta(F)
\end{array}
\right) q^{H+\bar H\pm 2};\\
\\
{[B,^tB]}=[N-\bar N],\ \  q^{N-\bar N}\left(\begin{array}{c}
B\\
^tB
\end{array}
\right)
= \left(\begin{array}{c}
B\\
^tB
\end{array}
\right) q^{N-\bar N\pm 2}.
\end{array}
\label{eq1.21}
\end{equation}
The subspace $\calh_I$ of $\calf\otimes \bar \calf$ which consist of
$U_q(sl_2)_\Delta\otimes U_q(sl_2)_{B,^tB}$ invariant vectors is spanned by
\begin{equation}
\begin{array}{c}
\{A^{+[n]}\vert \mbox{vac}\ \rangle,n=0,\dots,2h-2\},\  \ A^\pm=a^\pm_\beta
\bar a^\beta_\pm =A^\pm_1+A^\pm_2,\\
\\
 A^{[n]}=\frac{1}{[n]!}A^n=\displaystyle{\sum^{n-m}_{k=m}}
q^{k(n-k)}A_1^{[k]}A_2^{[n-k]},m=\max(0,n-h+1).
\end{array}
\label{eq1.22}
\end{equation}

\end{proposition}

A central result of \cite{FHT2} is the monodromy invariance of $g(x,t)$
(\ref{eq1.1}) when restricted to the physical subspace. It is reflected (within
the finite-dimensional zero-mode problem under consideration) in the property
\begin{equation}
a^\varepsilon M\bar M^{-1} \bar a_\varepsilon {\calh}_I=a^\varepsilon \bar
a_\varepsilon {\calh}_I
\label{eq1.23}
\end{equation}
which follows from (\ref{eq1.5}) and from
\begin{equation}
\left(q^{\pm(N-\bar N)}-1\right){\calh}_I=0.
\label{eq1.24}
\end{equation}

\Section{Generalized cohomologies in ${\calh}_I$}

The quadratic relations (\ref{eq1.6}) for $a^\pm_\alpha$ (and their
counterparts for $\bar a^\beta_\pm$) together with (\ref{eq1.3}) and
(\ref{eq1.8}) imply
\begin{equation}
[A,A^+]=[N+\bar N]=:[2\hat N],\ \ \ q^{\pm \hat N}A^\pm= A^\pm q^{\pm(\hat N\pm
1)}\ \ \mbox{for}\ \ A\equiv A^-.
\label{eq2.1}
\end{equation}
(Note that condition (\ref{eq1.24}) implies that the operator $\hat N$
introducted in (\ref{eq2.1}) has an integer spectrum on ${\calh}_I$.)

\begin{proposition}
Let $A^\pm_\alpha=a^\pm_\alpha \bar a^\alpha_\pm$, $\alpha=1,2$ (\underbar{no
summation}!); then (\ref{eq1.6}), (\ref{eq1.8}) and (\ref{eq1.16}) imply
\begin{equation}
A^\pm_2 A^\pm_1=q^2 A^\pm_1 A^\pm_2,\ \ (A^\pm_\alpha)^h=0\Longrightarrow
(A^\pm)^h=0
\label{eq2.2}
\end{equation}
\begin{equation}
A \vert n\rangle = [n]\vert n-1\rangle,n=1,\dots,2h-1\ \ ((\hat N-n)\vert
n\rangle=0, \vert 0 \rangle = 0)
\label{eq2.3}
\end{equation}
where $\vert n\rangle$ is defined by $\vert n\rangle := (A^+)^{[n-1]}\vert
\mbox{vac}\ \rangle$.
\end{proposition}
{\bf Proof}. The first equation (\ref{eq2.2}) follows from (\ref{eq1.8}) and
the $q$-Bose commutation relations
\begin{equation}
a^\pm_2 a^\pm_1 = q a^\pm_1 a^\pm_2,\ \ \bar a^2_\pm \bar a^1_\pm=q \bar
a^1_\pm \bar a^2_\pm
\label{eq2.4}
\end{equation}
derived from (\ref{eq1.6}). The second one is a direct consequence of
(\ref{eq1.16}). Finally, for $A^\pm=A^\pm_1+A^\pm_2$ we find
\begin{equation}
\begin{array}{ll}
A^n=\sum^n_{k=0} \left(\begin{array}{c}
n\\
k
\end{array}
\right)_+ A^k_1 A^{n-k}_2,\ &\mbox{where}\ \ \left(\begin{array}{c}
n\\
k
\end{array}
\right)_+ = \frac{(n)_+!}{(k)_+!(n-k)_+!},\\
\\
& \mbox{with}\ \ (n)_+ =\frac{q^{2n}-1}{q^2-1},
\end{array}
\label{eq2.5}
\end{equation}
which implies for $n=h$ the last equation (\ref{eq2.2}) since
\begin{equation}
(h)_+=q^{h-1}[h]=0\ \ \mbox{for}\ \ q^h=-1.
\label{eq2.6}
\end{equation}
For $n<h$,  (\ref{eq2.3}) follows directly from (\ref{eq1.22}) and
(\ref{eq2.1}). For $n\geq h$,  it follows from the relations
\[
A(A^+_1)^{[n]}\vert 1\rangle=q^{-1}[n](A^+_1)^{[n-1]}\vert 1\rangle,\ \
A(A^+_2)^{[n]}\vert 1\rangle=q[n](A^+_2)^{[n-1]}\vert 1\rangle
\]
($\vert 1\rangle\equiv\vert vac\ \rangle$) which allows to set in computing
(\ref{eq2.3})
\begin{equation}
(A^+_\alpha)^{[h]}=0,\ \ \alpha=1,2\ (\mbox{in}\ {\calh}_I).
\label{eq2.7}
\end{equation}
The relation
\begin{equation}
^t(A^+)=A\ \ (^tA=A^+)
\label{eq2.8}
\end{equation}
for the conjugation (\ref{eq1.18}) yields, as a corollary of Proposition 2.1,
the following Gram matrix (of inner products of the basis vectors)
\begin{equation}
\langle m\vert n\rangle = \langle 1 \vert A^{[m-1]}(A^+)^{[n-1]}\vert 1\rangle
=[n]\ \delta_{mn}.
\label{eq2.9}
\end{equation}
Noting that $[n]=-[h+n]\ (=[h-n])$ we deduce that the trace of this Gram matrix
is zero.\\
The space ${\calh}_I$ can be viewed as the complexification of the real span
${\calh}^{\mathbb R}_I$ of the vectors (\ref{eq1.22}). The fact that it admits
(unlike $\calf\otimes\bar\calf$) such a real basis (for which the Gram matrix
of inner products is real) allows to define a second, {\sl antilinear
anti-involution} on the operator algebra $\mathfrak A$  in ${\calh}_I$ defined
by
\begin{equation}
(A^\pm)^{[n]^+}=(A^\mp)^{[n]}
\label{eq2.10}
\end{equation}
Clearly, it coincides with the transposition $^t$ for $A^\pm$, but differs when
applied to $q^{\hat N}$~:
\begin{equation}
(q^{\hat N})^+=q^{-\hat N}\ \ (q^+=q^{-1}).
\label{eq2.11}
\end{equation}
(Note that the relation $q^{\hat N} A=A q^{\hat N-1}$ goes into $q^{\hat N
-1}A^+=A^+a^{\hat N}$ under transposition and into $q^{1-\hat N}
A^+=A^+q^{-\hat N}$ under the hermitian conjugation defined by (\ref{eq2.10})
and (\ref{eq2.11}); the full set of relations (\ref{eq2.1}) remains invariant
in both cases.). This allows to define an (indefinite) {\sl hermitian inner
product}
$\langle\bullet\vert\bullet\rangle_I$ in ${\calh}_I$ which coincides with the
real bilinear form $\langle\bullet\vert \bullet\rangle$ (inherited from
${\calf}\otimes\bar {\calf}$) on ${\calh}^\mathbb R_I$. In what follows we
shall only use this hermitian form and shall therefore drop the subscript
$I$.\\

Define the subsapce $\calh'$ of ${\calh}_I$, on which the hermitian form
$\langle\bullet\vert\bullet\rangle$ is positive semidefinite, by the set of
$h-1$ constraints
\begin{equation}
A^{h-1}(A^+)^n{\calh}'=0\ \ \mbox{for}\ \ n=0,1,\dots,h-2
\label{eq2.12}
\end{equation}
\begin{proposition}
The complement \mbox{Coim} $(A^{h-1}(A^+)^n)$ of the kernel of the operator
$A^{h-1}(A^+)^n$ in ${\calh}_I$ is 1-dimensional and given by
\begin{equation}
\mbox{Coim}\ (A^{h-1}(A^+)^n)=\{ \mathbb C \vert 2h-1-n\rangle \}
\label{eq2.13}
\end{equation}
\end{proposition}
The {\sl proof} is a straightforward consequence of (\ref{eq2.1}) and
(\ref{eq2.3}).
\\

We note that $A^\pm$ weakly commute with the constraints (\ref{eq2.12}). For
$A^+$ this follows from the easily verifiable relations
\[
[A^{h-1},A^+]=[2\hat N+h-2]A^{h-2}
\]
and
\[
[2\hat N+h-2]A^{h-2}{\calh}'=0.
\]
We also note that the operator $A^{h-1}(A^+)^{h-1}$ vanishes identically in
${\calh}_I$ and so does $A^+A^{h-1}$; both assertions follow from the identity
$A^+\vert h\rangle=0$.\\

Thus the space ${\calh}'$ is $h$-dimensional (spanned by $\vert n\rangle$ for
$1\leq n\leq h$). Its subsapce ${\calh}''$ of 0-norm vectors is 1-dimensional:
it consists of multiples of the vector $\vert h\rangle$.\\

The main result of this section (and of the paper) is the following realization
of the {\sl physical subquotient}
\begin{equation}
\calh ={\calh}'/{\calh}''
\label{eq2.14}
\end{equation}
in terms of generalized cohomologies of the type studied in \cite{D-VK} and
\cite{D-V}.
\begin{proposition}
Each of the generalized cohomologies of the nilpotent operator $A$ is
1-dimensional and can be realized as
\begin{equation}
H^{(n)}=\mbox{Ker}\  A^n/\mbox{Im}\  A^{h-n}\approx \{\mathbb C \vert
h-n\rangle\},\ \ n=1,\dots,h-1.
\label{eq2.15}
\end{equation}
With this choice the representative subspace is orthogonal to $\mbox{Im}\
A^{h-n}$. The physical subquotient (\ref{eq2.14}) is isomorphic to the direct
sum of $H^{(n)}$:
\begin{equation}
\calh \simeq \bigoplus^{h-1}_{n=1} H^{(n)}.
\label{eq2.16}
\end{equation}
The null space ${\calh}''$ in (\ref{eq2.14}) is isomorphic to the intersection
of images of $A^n$:
\begin{equation}
{\calh}''=\bigcap^{h-1}_{n=1} \mbox{Im}\  A^n =\mbox{Im}\  A^{h-1} (=\mbox{Im}\
 (A^+)^{h-1}).
\label{eq2.17}
\end{equation}
\end{proposition}
{\bf Proof}.\  $\mbox{Ker}A^n$\  is $2n$-dimensional and is spanned by the
vectors\linebreak[4] $\{\vert 1\rangle,\dots,\vert n\rangle; \vert
h\rangle,\dots,\vert h + n-1\rangle\}$. $\mbox{Im}\ A^{h-n}$ is
$(2n-1)$-dimensional and is spanned by the subset of the above, orthogonal to
$\vert h-n\rangle$. This proves (\ref{eq2.15}). The rest follows from the
explicit knowledge of $H^{(n)}$ and of $\mbox{Im}\ A^{h-n}$.

\Section{Discussion : Open problems}

The present cohomological treatment of a diagonal $SU(2)$ WZNW model leaves
open a number of related questions:
\begin{enumerate}
\item
The subspace ${\calh}'$ of ${\calf}\otimes\bar{\calf}$ (and hence the
corresponding subquotient $\calh$) was constructed in two steps. We first
singled out the subspace ${\calh}_I$ of ${\calf}\otimes\bar{\calf}$ of
$U_q(sl_2)\otimes U_q(sl_2)$ invariant vectors which admits a hermitian
(indefinite) inner product and then introduced the constraints that determine
${\calh}'$. The question whether one can introduce a (generalized) BRS charge
in ${\calf}\otimes\bar{\calf}$ that takes into account all constraints at the
same time is left open.
\item
Can non-diagonal $\hat{\mathfrak{su}}(2)$ models be treated in a similar
fashion ?
\item
Find the BRS cohomology of diagonal $\hat{\mathfrak{su}}(n)$ models for $n>2$;
their gauge theory treatment was initiated in \cite{FHT3}.
\end{enumerate}

\section*{Acknowledgments}

This work was initiated during the visits of one of the authors (I.T.) at the
La\-boratoire de Physique Th\'eorique et Hautes Energies, Universit\'e Paris
XI, Orsay and the Centre Emile Borel, Institut Henri Poincar\'e, Paris and was
completed during his stay in the Bogoliubov Laboratory of Theoretical Physics
of the Joint Institute for Nuclear Research, Dubna, Russia. I.T. would like to
thank all these institutions for hospitability and the Bulgarian National
Fondation for Scientific Research for partial support under contract F-404.

\end{document}